\documentclass[
    ,final            % use final for the camera ready runs
  ]
  {aipproc}

\layoutstyle{6x9}

\begin{document}

\title{AGN feedback in galaxy groups: a joint GMRT/X-ray study}


\author{S. Giacintucci}{
  address={Harvard-Smithsonian Center for Astrophysics, 60 Garden St, Cambridge, MA 02138, USA},altaddress={INAF-IRA, via Gobetti 
101, I-40129, Bologna, Italy}
}

\author{J. M. Vrtilek}{
  address={Harvard-Smithsonian Center for Astrophysics, 60 Garden St, Cambridge, MA 02138, USA}
}

\author{E. O'Sullivan}{
  address={Harvard-Smithsonian Center for Astrophysics, 60 Garden St, Cambridge, MA 02138, USA}
}

\author{S. Raychaudhury}{
  address={University of Birmingham, Edgbaston, Birmingham, UK}
}

\author{L. P. David}{
  address={Harvard-Smithsonian Center for Astrophysics, 60 Garden St, Cambridge, MA 02138, USA}
}

\author{T. Venturi}{
  address={INAF-IRA, via Gobetti 101, I-40129, Bologna, Italy}
}

\author{R. Athreya}{
  address={IISER, Pune, Maharashtra 411008, India India}
}

\author{M. Gitti}{
  address={Dept. of Astronomy, University of Bologna \& INAF OABo, via Ranzani 1, I-40127, Bologna, Italy}, altaddress={Harvard-Smithsonian Center for Astrophysics, 60 Garden St, Cambridge, MA 02138, USA}
}

\begin{abstract}
We present an ongoing study of 18 nearby galaxy groups, chosen for the 
availability of {\it Chandra} and/or {\it XMM-Newton} data and evidence 
for AGN/hot intragroup gas interaction. We have obtained 235 and 
610 MHz observations at the GMRT for all the groups, and 327 and 150 MHz 
for a few. We discuss two interesting cases -  NGC\,5044 and AWM\,4 - 
which exhibit different kinds of AGN/hot gas interaction. With the help of 
these examples we show how joining low-frequency radio data (to track the 
history of AGN outbursts through emission from aged electron 
populations) with X-ray data (to determine the state of hot gas, its 
disturbances, heating and cooling) can provide a unique insight into 
the nature of the feedback mechanism in galaxy groups.

\end{abstract}

\maketitle

%%%%%%%%%%%%%%%%%%%%%%%%%%%%%%%%%%%%%%%%%%%%
%% MAINMATTER
%%%%%%%%%%%%%%%%%%%%%%%%%%%%%%%%%%%%%%%%%%%%

\section{Introduction}

Several feedback mechanisms have been invoked to balance the cooling 
process in the hot X-ray emitting gas halo of clusters and groups of 
galaxies. These include gravitational heating, supernovae, subcluster mergers,
thermal conduction, and AGN-driven outbursts. In rich clusters, 
the radio outbursts from the AGN harboured in the central galaxy appear to be 
the dominant source of heating. However most of the galaxies and baryonic 
matter in the Universe reside in less massive units, such as poor clusters 
and groups of galaxies. Thus the study of feedback in these environments is 
fundamental to understand the mechamism by which feedback operates and how 
it has influenced the formation and thermal history of most of the baryons in 
the Universe.

\section{Our ongoing GMRT/X-ray study of galaxy groups} 

To investigate the impact of AGN feedback in galaxy groups, we are carrying 
out an in-depth study of a representative collection of 18 galaxy 
groups (listed in Tab.~1), using low (and multi) frequency radio observations, 
obtained at the Giant Metrewave Radio Telescope (GMRT), and deep archival 
{\it Chandra} and {\it XMM-Newton} X-ray data. The groups were chosen
on the basis of structures, either in the X-ray surface brightness and 
temperature maps or radio morphology, which strongly indicate interaction 
between the radio source and the intragroup medium. All groups, 
except two, were observed with the GMRT both at 610 MHz and 235 MHz during 
Cycle 12, 14 and 15. The observing period for each frequency is given in 
Tab.~1. GMRT observations at 327 MHz and 150 MHz were also obtained for
few targets (see notes to Tab.~1) as part of the follow-up ongoing 
at these frequencies. The data from all observations in Tab.~1 have been 
completely reduced. The sensitivity achieved in the final images is in 
the range 35-100 $\mu$Jy at 610 MHz and 0.2-1 mJy at 235 MHz for a 
typical observing time of $\sim$2-3 hours on source. The ongoing analysis 
shows that the sources are diverse, covering a range of spatial scales 
and total powers, from classic double FR-I to core-halo radio sources 
(Giacintucci et al. in preparation). 

\begin{table}
\begin{tabular}{lccclcc}
\hline
\tablehead{1}{r}{b}{Group name} & \tablehead{1}{r}{b}{235 MHz}
  & \tablehead{1}{r}{b}{610 MHz} &  \tablehead{1}{r}{b}{~}& \tablehead{1}{r}{b}{Group name} &
\tablehead{1}{r}{b}{235 MHz}  & \tablehead{1}{r}{b}{610 MHz}\\

\hline

UGC\,408\tablenote{observed at 150 MHz in Cycle 16 (summer 2009)} & Aug 2007 & Aug 2008 
& &NGC\,3411 & Aug 2006 & Feb 2008 \\
NGC\,315 & Feb 2008 & Aug 2008 & &NGC\,4636 & Aug 2006 & Feb 2008 \\
NGC\,383 & Feb 2008 & Aug 2008 & &HCG\,62   & Feb 2008 & Feb 2008 \\
NGC\,507 & July 2006 & Aug 2008 & &NGC\,5044$^{*} \, $\tablenote{observations presented 
in David et al. (2009); also observed at 327 MHz;} & 
Feb 2008 & Feb 2008 \\
NGC\,741 & Aug 2006 & Aug 2007 & &NGC\,5813$^{*}$ & $-$ & Aug 2008 \\
HCG\,15  & Aug 2006 & Aug 2008 & &NGC\,5846 & Aug 2006 & $-$\\   
NGC\,1407& July 2006 & Aug 2008 & &AWM\,4 \tablenote{observations presented in Giacintucci 
et al. (2008); also observed at 327 MHz;}& Aug 2006 & Jul 2006 \\
NGC\,1587& Aug 2006 & Aug 2008 & & NGC\,6269  & Feb 2008 & Feb 2008 \\
MKW\,02\tablenote{observations from Giacintucci et al. (2007).}  & Aug 2003 & July 2005& & 
NGC\,7626 $^{*}$ & Aug 2007 & Aug 2008 \\ 

%1998\tablenote{predicted}& & 200 & 300 & 1500  & 2000\\
\hline
\end{tabular}
\caption{List of galaxy groups and status of the GMRT observations.}
\label{tab:a}
\end{table}

\section{A closer view of two remarkable galaxy groups}

We have selected two systems from the above sample - NGC\,5044 and AWM\,4 - 
which exhibit different kinds of AGN/hot gas interaction, to show how the 
combination of deep X-ray data and high-sensitivity low frequency radio 
observations can provide a unique insight into the nature of the feedback 
mechanism in galaxy groups.

\paragraph{{\bf NGC\,5044}} The new {\it Chandra} observation reveals that 
the group hosts many small radio quiet cavities, filaments, and a semi-circular 
cold front (Fig.~1; David et al. 2009). The GMRT 610 MHz image (right) shows 
the radio core and a lobe which extends along a filament of cold 
gas. The 235 MHz emission (left) is much more extended, with little overlap 
with the 610 MHz image. The emission fills the S cavity, curves 
toward the W just behind the cold front, and then sharply bends by 
$\sim$90$^{\circ}$. A second component, a detached radio lobe, is visible 
toward the S-E, and its western edge is coincident with the cold front, 
suggesting that the relativistic material in this structure might have been 
produced by an earlier outburst and that it is currently compressed by the 
motion of NGC\,5044 toward the S-E. The lack of any detected emission at 610 
MHz in the same region as the 235 MHz emission indicates that the radio 
spectrum must be very steep. The GMRT data thus appear to reveal 2 (or 
possibly 3) separate outbursts. The youngest outburst can be identified with 
the 610 MHz emission, and the oldest burst with the detached radio lobe.

\begin{figure}
  \includegraphics[height=.29\textheight]{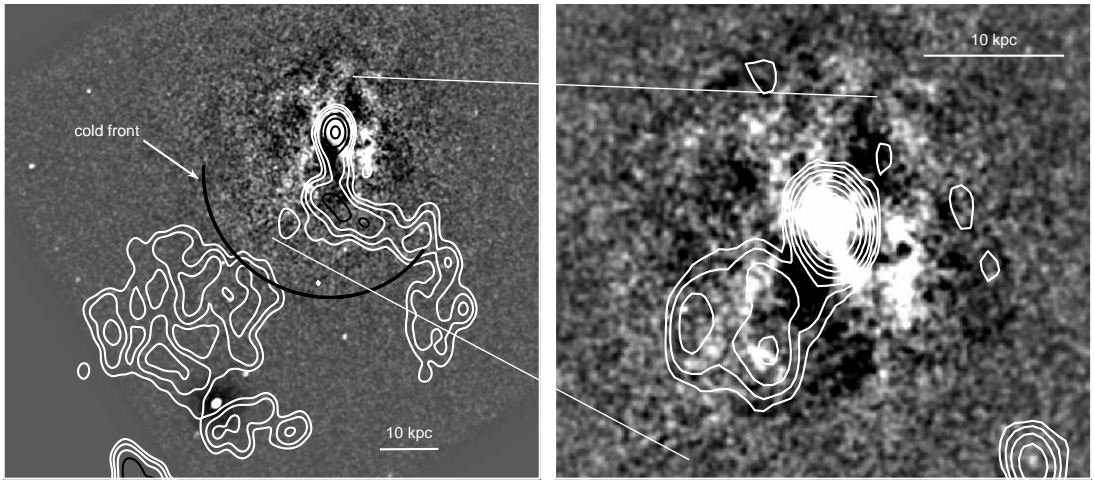}
  \caption{NGC\,5044: GMRT 235 MHz (left) and 610 MHz contours (right)
on the {\it Chandra} unsharp masked image in the 0.3-2.0 keV band 
(David et al. 2009). The radio beam is 22$^{\prime \prime} \times 
16^{\prime \prime}$ and 18$^{\prime \prime} \times 16^{\prime \prime}$, 
respectively. The lowest contour is shown at 3${\sigma}$=0.75 mJy 
b$^{-1}$ and  0.075 mJy b$^{-1}$, respectively.}
\end{figure}

\begin{figure}
  \includegraphics[height=.29\textheight]{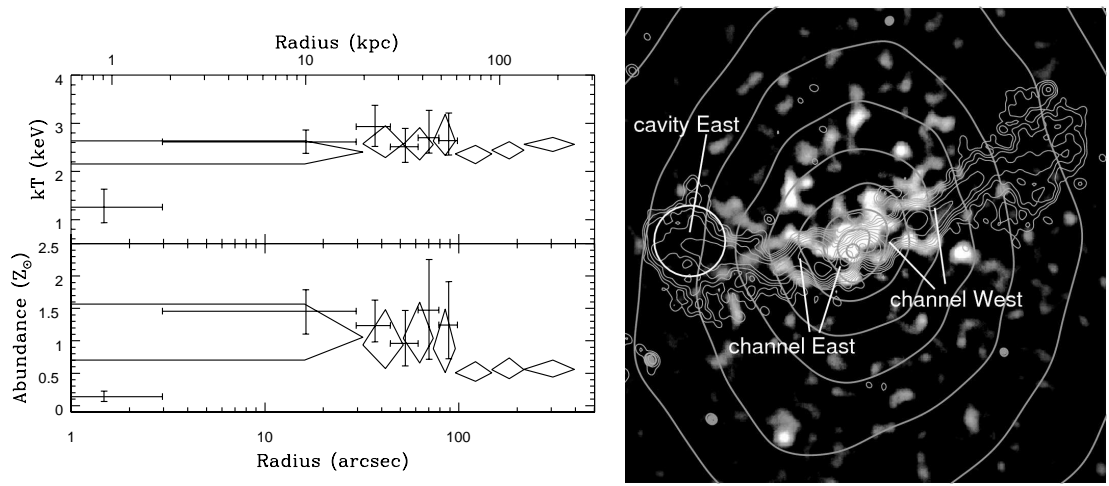}
  \caption{AWM\,4. {\it Left:} {\it Chandra} radial profiles of deprojected 
temperature and abundance. Crosses and diamonds refer to fits carried out 
in the 0.7-7.0 keV band with 8000 net counts/spectrum (370 cts for the 
central bin) and 12000 net counts/spectrum, respectively. {\it Right}: 
GMRT 610 MHz contours (beam 5$^{\prime \prime} \times 4^{\prime \prime}$; 
lowest contour at 3${\sigma}$=0.15 mJy b$^{-1}$), superposed on the 
{\it Chandra} residual image, after subtraction of the best-fitting surface 
brightness model.}
\end{figure}

\paragraph{{\bf AWM\,4}} Previous XMM-Newton observations showed AWM\,4 
to be isothermal at $\sim$2.5 keV out to at least 160 kpc from the centre, 
even though the cooling time in the middle is $\sim$ 2 Gyr (O'Sullivan et 
al. 2005). Its powerful central radio galaxy 
was proposed as the 
most likely source of heating. Our deep GMRT observations at 235, 327 
and 610 MHz, presented in 
Giacintucci et al. (2008), revealed the full extent of the radio 
source and allowed us to determine its age, orientation, energy and 
physical parameters. However, with no indications of X-ray cavities 
associated with the radio source, the question of the coupling 
between jets and intra-group gas remained 
unresolved (see also Gastaldello et al. 2008). Our new 80 ksec {\it Chandra} image reveals the 
small-scale galactic corona surrounding the 
AGN; the radial temperature and abundance profiles show a clear 
decrease in the central $\sim$2 kpc-radius 
region (Fig.~2; left), corresponding to the corona. 
The existence of this corona might exlain the long timescale of 
the outburst, as it fuels the AGN but is largely unaffected by the 
radio jets. We do not detect clear 
cavities in the {\it Chandra} image associated with the radio lobes. 
Some weak X-ray features are visible in 
the residual image obtained after subtraction of the best-fitting 
surface brightness model, shown in 
Fig.~2 (right) with overlaid the GMRT 610 MHz contours. A faint 
depression is visible in the region of the
eastern lobe, and two channel--like structures appear to be associated 
with the inner jets. No cavity is 
detected in the western lobe. The detailed analysis of these features 
and their implications is ongoing (O'Sullivan et al. in prep.).

\section{Summary}

Our ongoing X-ray/GMRT study of the AGN feedback in galaxy groups 
shows that low frequency 
observations play a crucial role in the estimate of the total radio 
energy input from the central 
AGN and are important in the study of the history of the energy 
injection and transfer, since they
can reveal old, steep-spectrum radio emission and structures 
related to previous AGN outbursts (e.g., 
NGC\,5044). As shown by our analysis of AWM\,4, the combination of 
the low frequency information with 
deep and high resolution X-ray data offers a unique tool to 
investigate the AGN/hot gas interaction
(see also Gitti et al., these proceedings). This is of fundamental 
importance to determine the 
thermal history of the intragroup and intracluster medium and 
shed light on the role of the AGN feedback 
on the formation and evolution of galaxies in these environments.

%%%%%%%%%%%%%%%%%%%%%%%%%%%%%%%%%%%%%%%%%%%%%%%%
%% BACKMATTER
%%%%%%%%%%%%%%%%%%%%%%%%%%%%%%%%%%%%%%%%%%%%%%%%

\begin{theacknowledgments}
We thank the staff of the GMRT for their help during 
the observations. GMRT is run by the National Centre for Radio Astrophysics 
of the Tata Institute of Fundamental Research. We thank C. Jones, W. Forman, 
P. Mazzotta, M. Murgia, and T. Ponman.

\end{theacknowledgments}

\end{document}